\documentstyle[preprint,aps,floats,fleqn]{revtex}

\ifx\nopictures Y \else \input epsf.tex \fi
\tighten
\begin{document}

\def\pAl{p_{A, l}}                               
\def\pBl{p_{B, l}}
\def\pal{p_{a, l}}                                 
\def\pbl{p_{b, l}}

\def\pAc{p_{A, cm}}                             
\def\pBc{p_{B, cm}}
\def\pac{p_{a, cm}}                             
\def\pbc{p_{b, cm}}

\def\pAh{p_{A, h}}                             
\def\pBh{p_{B, h}}
\def\pah{p_{a, h}}                             
\def\pbh{p_{b, h}}

\def\xh{\widehat x}
\def\zh{\widehat z}
\def\qTh{\widehat{q_T}}
\def\sh{\widehat s}
\def\th{\widehat t}
\def\uh{\widehat u}
\def\Wh{\widehat W}
\def\Vh{\widehat{V}}
\def \sigh {\widehat{\sigma}}

\def\beq{\begin{equation}}
\def\eeq{\end{equation}}
\def\bdm{\begin{displaymath}}
\def\edm{\end{displaymath}}
\def\bea{\begin{eqnarray}}
\def\eea{\end{eqnarray}}

\def\lp{\l^{\prime}}
\def\SeA{S_{eA}}                                        
\def\vg{\gamma^{*}}                                     
\def\alpi{\frac{\alpha_s}{\pi}}                         
\def\sFs{\frac{\sigma_0 F_l}{\SeA}}

\def\D0{D\0~}
\def\ov{\overline}
\def\ra{\rightarrow}
\def\cms{c.m. }
\def\MSbar{$\overline{MS}$ }

\def\mupr{\mu^{\prime}}
\def\eps{\varepsilon}
\newcommand{\prescr}[2]{{\,}^{#1}{#2}}
\def\Mvariable{}

\setcounter{footnote}{1}
\renewcommand{\thefootnote}{\fnsymbol{footnote}}

\preprint{\small \tighten 
\parbox{5.5cm}{MSUHEP-01213\\ CTEQ-016\\hep-ph/0012261\\}
}

\title{
Phenomenology of multiple parton radiation in semi-inclusive
deep-inelastic scattering
}

\author{ P.M. Nadolsky\thanks{E-mail address: nadolsky@pa.msu.edu}, 
         D. R. Stump\thanks{E-mail address: stump@pa.msu.edu}
	and C.-P. Yuan \thanks{E-mail address: yuan@pa.msu.edu}}
\address{\rm Department of Physics and Astronomy,
Michigan State
University,\\ East Lansing, MI 48824, USA}

\date{\today}
\maketitle
\begin{abstract}
\tighten
In the current region of semi-inclusive deep-inelastic
scattering $\l + A \ra l^\prime + B +X$, most events are
accompanied by intensive radiation of soft and collinear partons which
cannot be reliably described at any fixed order of perturbative
QCD. In this paper, a resummation formalism that 
describes such multiple parton radiation is compared
to the HERA data on semi-inclusive DIS, 
including the distributions of energy flow and charged
particle multiplicity.
We show that the resummation of multiple parton radiation improves 
the agreement
between the theory and the data. We 
make some suggestions on further experimental
study of multiple parton radiation at HERA.
\end{abstract}
\newpage

\widetext

\section {Introduction}

In a previous paper \cite{nsy1999}, we presented
a formalism that enhances the predictive power of perturbative
Quantum Chromodynamics (pQCD) with respect to the description of
semi-inclusive hadroproduction
at lepton-hadron colliders. We continue that study here,
in part because new data has become available from HERA.
The reaction that we consider is  
$l + A \ra l^\prime + B + X$,
where $l, A$ and $l^\prime,  B$ are initial- and final-state
leptons and hadrons, respectively. 
Calculations in low orders of pQCD are only partially
successful in explaining the abundant
experimental data on semi-inclusive 
deep-inelastic scattering (sDIS)
produced in recent years at the DESY-HERA
$ep$ collider~ \cite{H1chgd,H1z1,H1z2,ZEUSchgd96,ZEUSasym}.  Satisfactory 
agreement between the theory predictions and the
experimental data requires 
scrupulous and systematic calculation of
higher-order radiative corrections and, when necessary,
all-order summation of poorly converging perturbative series. A formalism
that was originally proposed in \cite{Meng2} and
extended in \cite{nsy1999} serves the latter purpose. 
It provides an improved theoretical description 
of nearly massless final-state hadrons that escape near
the direction of the momentum of the electroweak vector boson
exchanged between the leptonic and hadronic parts of the sDIS process.
In other words,
this formalism describes production of hadrons 
in the region of current fragmentation. 
In this kinematical regime, multiple radiation of soft and collinear
partons is a dominant factor influencing the behavior of
experimental observables. We have shown that 
the effects of multiple parton radiation
are most evident in pseudorapidity spectra 
in the center-of-mass frame of the initial hadron and the
exchanged electroweak boson.

The main goal of this paper is to demonstrate that resummation of
large logarithmic corrections in the current region of sDIS 
improves the theoretical description of many
aspects of this process. We compare predictions made within  the
resummation formalism to various sets of experimental data and to
predictions obtained for these observables from a next-to-leading order (NLO)
perturbative QCD calculation.
 
In Section\,\ref{SecSummary}, 
we present the equations
for the ${\cal O}(\alpha_s)$ and resummed cross sections
that will be compared to experimental data in
the subsequent Sections. In Section\,\ref{SecET}, 
we compare theoretical predictions to new HERA data \cite{H1z2} on
the energy flow in the hadronic center-of-mass 
reference frame. In Section\,\ref{SecChgd},
we discuss the impact of multiple parton radiation on various aspects
of semi-inclusive
charged particle production: in particular, resummation
effects on transverse momentum distributions and on the 
correlation between the transverse and longitudinal momentum components of
the observed final-state hadron. Section\,\ref{SecConclusion}
contains our conclusions.

\section{Summary of the fixed-order and 
resummation calculations}
\label{SecSummary}
In this paper we use notations and variables that
were originally defined in Sections\,II and III of \cite{nsy1999}.
For the convenience of the reader, we briefly list the relevant definitions
in Appendix\,{A}.

Following \cite{Meng1,Meng2}, we introduce a momentum scale $q_T$ 
related to the space-like vector $q_t^{\mu}$ given by
\beq
q_{t}^{\mu}=q^{\mu}-p_{A}^{\mu}\frac{q\cdot p_{B}}{p_{A}\cdot p_{B}}
-p_{B}^{\mu}\frac{q\cdot p_{A}}{p_{A}\cdot p_{B}}. \label{qTmu}
\eeq
The square of the scale $q_T$ is defined by
\beq
q_{T}^{2}=-q_{t}^{\mu}q_{t\mu}.
\label{qT2}
\eeq
Throughout this paper we will consider observables in the 
hadronic center-of-mass (hCM) frame, in which 
the four-vector $q_t^\mu$ has a simple interpretation.
This frame is defined by the condition $\vec p_A + \vec q = 0$.
Let the direction of the $z$-axis in
the hCM frame coincide with the direction of $\vec q$. Then
the energy component and $z$ component of $q_t^\mu$ vanish. The   
transverse components form a two-dimensional vector 
$\vec q_T$ that is simply the transverse momentum $\vec p_T$
of the final hadron $B$
rescaled by the Lorentz scalar $-z\equiv\,-(p_{B}\cdot
p_{A})/(q\cdot p_{A})$; that is, 
\beq
q_{t}^\mu = \{0,\vec q_T, 0\} = \{0,-\frac{\vec p_T}{z}, 0\}.
\eeq
The variable $z$ controls the fragmentation of partons in the final state. 
Thus the vector $\vec q_T$ is closely related to the 
transverse momentum $\vec p_T$, 
except that it is not sensitive to
the magnitude of the momentum vector $\vec p_B$, {\it i.e.,} not sensitive to 
the collinear final-state fragmentation.
Rather, $\vec q_T$ determines the {\em direction} in which the 
final-state hadron is produced. 
In both the hadron Breit frame and the hCM frame, the magnitude of this
vector $q_T$
is a function of the pseudorapidity of the final-state hadron.
In particular, we will later refer to the relationship between $q_T$ and
the pseudorapidity $\eta^{cm}$
of the observed hadronic final state in the hCM frame, which is 
\beq
q_T = W e^{-\eta^{cm}},
\label{qTeta}
\eeq
where $W^2 =( p_A +q )^2 = Q^2 (1/x -1)$ is the square
of the center-of-mass energy of the virtual photon and the proton. 
The multiple parton radiation that
we are discussing here is important in the limit $\eta^{cm} \ra
+\infty$, or, equivalently,  $q_T \ra 0$.

To see how convergence of the perturbative series
breaks down in the small-$q_T$ region, we start by examining
the hadron-level sDIS cross section
$d\sigma_{BA}/(dx dz dQ^{2} dq_{T}^{2} d\varphi)$. This cross section can be
expressed in terms of convolutions of  parton-level cross sections 
$d\widehat \sigma_{ba}/(d\widehat{x} d\widehat{z} dQ^{2} dq_{T}^{2} d\varphi)$,
and phenomenological long-distance hadronic functions: 
\beq
\frac{d\sigma_{BA}}{dx dz dQ^{2} dq_{T}^{2} d\varphi}=\sum_{a,b}
\int_{z}^{1}
\frac{d\xi_{b}}{\xi_{b}} D_{B/b}(\xi_{b}, \mu_F)\int_{x}^{1}
\frac{d\xi_{a}}{\xi_{a}}
F_{a/A}(\xi_{a}, \mu_F)
\frac{d\widehat \sigma_{ba}(\mu_F)}{d\widehat{x} d\widehat{z} dQ^{2} dq_{T}^{2} d\varphi}.
\label{factcs}
\eeq
Here $F_{a/A}(\xi_{a}, \mu_F)$ denotes the parton distribution function (PDF)
for a parton $a$ in the hadron $A$,  and
$D_{B/b}(\xi_{b}, \mu_F)$ is the fragmentation function (FF) for the hadron
$B$ that is produced in the fragmentation of a parton $b$. 
The notation $\mu_F$ is used for the scale of factorization. 
The sum over the labels $a, b$ includes contributions from all parton types,
{\it i.e.,} $g, u, \bar u, d, \bar d, \dots\ $. In the
following, a sum over the indices $i,j$ will include contributions
from active flavors of quarks and antiquarks only, {\it i.e.,} it will not
include a gluonic contribution.

Both the hadron-level and parton-level cross sections can be expanded 
into sums over leptonic angular functions $A_\rho (\psi, \varphi)$ (which are listed in Appendix A), as
\bea
\frac{d \sigma_{BA}}{d x d z  dQ^{2}  dq_T^{2} d\varphi} &=&
\sum_{\rho=1}^{4} 
\prescr{\rho}{V}_{BA}(x,z,Q^2,q_T^2)
A_{\rho}(\psi,\varphi),
\label{Ang1} \\
\frac{d \sigh_{ba}}{d \xh d \zh  dQ^{2}  dq_T^{2} d\varphi} &=&
\sum_{\rho=1}^{4} 
\prescr{\rho}{\Vh}_{ba}(\xh,\zh,Q^2,q_T^2)
A_{\rho}(\psi,\varphi).
\label{ang1}
\eea
The coefficients
$\prescr{\rho}{V_{BA}}$ (or $\prescr{\rho}{\Vh_{ba}}$)
of the angular functions $A_{\rho}(\psi,\varphi)$ are independent of
one another.
If $q_T/Q \ll 1$, but $q_T \neq 0$, the perturbative parton-level
cross section 
is dominated by the term with $\rho = 1$.
The structure function for this term
$ \prescr{1}{\Vh_{ba}}(\xh, \zh, Q^2, q_T^2)$ behaves
as $1/q_T^2$ times a series in powers of $\alpha_s$ and logarithms 
$\ln (q_T^2/Q^2$),
\beq
\prescr{1}{\Vh}_{ba}(\xh, \zh, Q^2, q_T^2) \approx
\frac{\sigma_0 F_l}{2 \pi \SeA} 
\frac{1}{q_T^2} \sum_{k=1}^{\infty} 
\Biggl(\frac{\alpha_s}{\pi}\Biggr)^k \sum_{m = 0}^{2 k - 1} 
\widehat v^{(km)}_{b a}(\xh,\zh) \ln^m \Biggl( \frac{q_T^2}{Q^2}\Biggr).
\label{NLL}
\eeq
The constants $\sigma_0$ and $F_l$, which were 
introduced in \cite{nsy1999}, are
\beq
\sigma_0 \equiv \frac{Q^2}{ 4 \pi \SeA x^2} \Bigl( \frac{e^2}{2} \Bigr),
\eeq
\beq
F_l = \frac{e^2}{2} \frac{1}{Q^2}.
\eeq
The coefficients $\widehat v^{(km)}_{b a}(\xh,\zh)$ 
are generalized functions of the variables $\xh$ and $\zh$.

Convergence of the series in (\ref{NLL}) deteriorates rapidly as
$q_T/Q \ra 0$, because of the growth of the
terms $(q_T^{-2})\ln^m (q_T^2/Q^2)$. As a result,  
the structure function $ \prescr{1}{\Vh_{ba}}(\xh, \zh, Q^2, q_T^2)$
has a non-integrable singularity at $q_T=0$. Its asymptotic behavior is very
different
from that of the structure functions $ \prescr{2,3,4}{\Vh_{ba}}(\xh, \zh,
Q^2, q_T^2)$, which are less singular
and, in fact, integrable at $q_T = 0$.
This singular behavior of 
$ \prescr{1}{\Vh_{ba}}(\xh, \zh, Q^2, q_T^2)$ is generated 
by infrared singularities of the perturbative cross section that
are located at $q_T=0$. These singularities 
must be either explicitly canceled or factored
into the long-distance hadronic functions in order to obtain a finite
cross section. 

There exist two qualitatively different approaches for handling such
singularities. The first approach deals with the singularities 
order by order in perturbation theory; 
the second approach identifies and sums the most singular 
terms in all orders of the perturbative expansion. 
We will begin by discussing the first approach, in which
singularities are regularized independently at each
order of the series in $\alpha_s$. In the crudest approximation, one 
keeps only terms of the order ${\cal O}(\alpha_s)$ and drops all
higher-order
terms. The singularity in
the ${\cal O}(\alpha_s)$ part of the asymptotic
expansion (\ref{NLL}) can be regularized by introducing a ``separation  
scale'' $q_T^{sep}$ and considering 
the fixed-order cross section separately
in the regions $0 \leq q_T \leq q_T^{sep}$ and $q_T > q_T^{sep}$.
The value of $q_T^{sep}$ should be
small enough for the approximation (\ref{NLL}) to
be valid over the whole range $q_T \leq q_T^{sep}$.

In the region $0 \leq q_T \leq q_T^{sep}$, we can apply the \MSbar
factorization scheme to take care of the singularities at $q_T = 0$.
In the \MSbar  scheme, the regularization is done through
continuation of the   
parton-level cross section to $n=4-2\eps$ dimensions. 
The $n$-dimensional expression for the asymptotic expansion (\ref{NLL})
of
$\prescr{1}{\Vh}_{ba}(\xh, \zh, Q^2, q_T^2)$ is
\beq
\prescr{1}{\Vh}_{ba}(\xh, \zh, Q^2, q_T^2) \approx
\frac{\sigma_0 F_l}{2 \pi \SeA}
\Biggl(\frac{2 \pi \mu}{\zh}\Biggr)^{4-n} 
\frac{1}{q_T^2} \sum_{k=1}^{\infty} 
\Biggl(\frac{\alpha_s}{\pi}\Biggr)^k \sum_{m = 0}^{2 k - 1} 
\widehat v^{(km)}_{b a}(\xh,\zh) \ln^m \Biggl( \frac{q_T^2}{Q^2}\Biggr).
\label{NLL2}
\eeq
The scale parameter $\mu$ 
in (\ref{NLL2}) is introduced to restore the correct
dimensionality of the parton-level cross section 
$d\widehat \sigma_{ba}/(d\widehat{x} d\widehat{z} dQ^{2} dq_{T}^{n-2} d\varphi)$. 
The soft and collinear singularities 
appear 
as terms proportional to $1/\eps^n$ when $n\ra 4$.
The soft singularity in the real emission corrections 
cancels with the soft singularity in the virtual corrections. 
The remaining collinear singularities
are absorbed into the partonic PDFs and FFs, 
which therefore 
must be redefined at each order of the perturbative calculation.
When the partonic PDFs and FFs are subtracted from 
the partonic cross section $d \widehat\sigma$,
the remainder 
is finite and independent of the types of the external hadrons.
We denote this finite remainder as
$(d \widehat\sigma)_{hard}$. The convolution of $(d
\widehat\sigma)_{hard}$ with the hadronic
PDFs and FFs yields
a cross section for the external hadronic states $A$ and $B$.
There is a certain freedom in the precise definition of 
the collinear contributions, so that the ``hard'' part   
depends on an arbitrary factorization scale $\mu_F$ through 
terms like $P_{ab} \ln (\mu_F/K)$, where $P_{ab}(\xi)$  are
${\cal O}(\alpha_s)$ splitting functions \cite{DGLAP}, 
and $K$ is some momentum scale in the process.\footnote{
The scales $\mu_F$ and $\mu$ are related as 
$\mu_F^2 = 4\pi e^{-\gamma_E} \mu^2 $, where $\gamma_E$ 
is the Euler constant.}
The dependence on the factorization scale in
the hard part is compensated, up to higher-order terms in
$\alpha_s$, by scale dependence of the long-distance hadronic  
functions.\footnote{In our calculations we do not distinguish between 
factorization scales for PDFs and FFs nor between the
factorization and renormalization scales.}

After the cancellation of soft singularities and factorization of collinear
singularities, 
one can calculate analytically the integral of  
$(\prescr{1}{\Vh_{ba}})_{hard}$ over the region 
$0 \leq q_T \leq
q_T^{sep}$.  At ${\cal
O}(\alpha_s)$ this integral is given by  
\beq
\int_{0}^{(q_T^{sep})^2} d q_T^2 \left(\prescr{1}{\Vh_{ba}}\right)_{hard} = 
\sFs \sum_{j} e_j^2 
\Biggl\{\prescr{1}{\Vh^{LO}_{ba,j}} + \frac{\alpha_s(\mu_F^\prime)}{\pi}
\prescr{1}{\Vh^{NLO}_{ba,j}} \Biggr\},  
\label{NLOcs}
\eeq
where $j=u,\bar u, d,\bar d,\dots\ $; the LO and NLO
structure functions are  
\bea
\prescr{1}{\Vh^{LO}_{ba,j}}   &=& \delta(1-\zh) \delta(1-\xh)
\delta_{b j}\delta_{j a},         \\
\prescr{1}{\Vh^{NLO}_{ba,j}}  &=&  - \frac{1}{2}
    \Biggl[
       \Biggl(C_F \ln^2{ \frac{Q^2}{(q_T^{sep})^2}} - 3 C_F
      \ln{\frac{Q^2}{(q_T^{sep})^2}}  \Biggr) 
        \delta(1-\zh) \delta(1-\xh) \delta_{b j}
        \delta_{j a}   
                             \nonumber \\
 &  + & \ln{\frac{\mu_F^2}{(\zh \,q_T^{sep})^2}}\biggl(
       \delta(1-\zh) \delta_{b j} P_{j a} (\xh)  
      + P_{b j} (\zh)\delta(1-\xh)\delta_{j a} 
       \biggr)\Biggr]      \nonumber           \\
 & + & \delta(1-\zh) 
    \delta_{b j}
     {c}^{in (1) }_{ja}(\xh)   +    
     {c}^{out (1) }_{bj } (\zh) \delta(1-\xh)\delta_{ja}. 
\label{V1NLO}
\eea
The coefficient functions ${c}^{in,out (1)}_{ba}(\xi)$ that appear in
$\prescr{1}{\Vh^{NLO}_{ba,j}}$ are given by
\bea
 {c}^{(1) in}_{j i} (\xi)= {c}^{out (1)}_{ij} (\xi) &=&
\delta_{ij}C_F \Biggl[\frac{1}{2} (1-\xi)  
- 2 \delta ( 1- \xi) \Biggr], \\
 {c}^{(1) in}_{j g} (\xi) &=& \frac{1}{2} \xi (1-\xi), \\
 {c}^{(1) out}_{g j} (\xi)&=& \frac{C_F}{2} \xi.
\label{Csdis}
\eea  

Now consider the kinematical region $q_T > q_T^{sep}$, where
the approximation (\ref{NLL2})  no longer holds.
In this region,  $(\prescr{1}{\Vh_{ba}})_{hard}$ should be obtained 
from the exact NLO result.
With this prescription, the
integral over $q_T^2$
can be calculated as 
\bea
\int_{0}^{\max q_T^2} d q_T^2 
\frac{d \widehat \sigma_{ba}}{d \xh d\zh d Q^2 dq_T^2 d\varphi}  
&=& \nonumber \\
A_{1}(\psi,\varphi)
\Biggl\{\int_{0}^{(q_T^{sep})^2} d q_T^2 
\left(\prescr{1}{\Vh_{ba}}\right)_{hard}
&+& \int_{(q_T^{sep})^2}^{\max q_T^2} d q_T^2 
\left(\prescr{1}{\Vh_{ba}}\right)_{hard}
\Biggr\} \nonumber \\
&+& \sum_{\rho=2}^{4} A_{\rho}(\psi,\varphi)\
\int_{0}^{\max q_T^2} d q_T^2 
\left(\prescr{\rho}{\Vh_{ba}}\right),
\label{pert}
\eea
where $\max q_T^2$ is the maximal value of $q_T^2$ allowed by
kinematics. The first integral on the right-hand side is calculated 
analytically, using the
approximation (\ref{NLOcs}); the second and third integrals are
calculated numerically, 
using the complete perturbative result of the order ${\cal
O}(\alpha_s)$.  The numerical calculation is done with the help of a
Monte Carlo integration program written in the style of the 
package ResBos used earlier
for resummation in vector boson production at hadron-hadron
colliders \cite{BY}.

A significant failure 
of the calculational procedure in (\ref{pert}) is that it
does not reproduce the correct shape of $q_T$-dependent 
differential cross sections in the
region $q_T \ll Q$. The fundamental problem is that the terms 
in (\ref{NLL2}) with
small powers of $\alpha_s$  do not reliably approximate
the complete sum in the region $q_T \ll Q$, where the large magnitudes of the
logarithmic terms $\ln^m (q_T^2/Q^2)$ quickly invalidate the expansion in 
powers of $\alpha_s$. In other words, a truncated sum 
in powers of $\alpha_s$ ({\it cf.} Eq.\,(\ref{NLOcs})) 
cannot be a reliable prediction 
for the observable cross section in the
region $q_T/Q \ll 1$.

The problem mentioned in the previous paragraph 
justifies the second approach to the regularization of the
 singularities at $q_T = 0$, in which  
 large logarithms in (\ref{NLL2}) are summed to all orders.
According to the formalism developed in \cite{CSepem,CSS} and adapted to the
 sDIS process in \cite{Meng2}, a
 better approximation for this sum is provided by
the Fourier transform of a $\vec b$-space function 
$\widetilde W(b,Q, \widehat x, \widehat z, C_1,C_2, \mu_F)$, 
which sums the dominant terms in (\ref{NLL}) through all orders of $\alpha_s$:
\begin{equation}
\label{resum}
\prescr{1}{\Vh}_{ba}(\xh, \zh, Q^2, q_T^2) =
\frac{\sigma_0 F_l}{2 S_{eA}}
\int \frac{d^{2}b}{(2\pi)^{2}} e^{i\vec q_T \cdot \vec
b } \widetilde W_{ba}(b, Q, \xh, \zh , C_1, C_2, \mu_F).
\end{equation}
Here $\vec b$ is a vector conjugate to $\vec q_T$, and $b$ denotes
the magnitude of $\vec b$.  

From first principles \cite{CSepem,CSS}, 
the general structure of $\widetilde W_{ba}(b,Q,\xh,\zh,C_1,C_2,\mu_F)$ 
is known  to have the form
\bdm
\widetilde W_{ba}(b,Q,\widehat x, \widehat z, C_1,C_2,\mu_F) = 
\edm
\beq
\sum_{j} e_j^2 
{\cal C}^{out}_{bj}(\widehat z,C_1,C_2,b_*,\mu_F) 
e^{-S^P(b_*,Q,C_1,C_2)}e^{-S^{NP}} {\cal C}_{ja}^{in} (\widehat x,C_1,C_2,b_*,\mu_F),
\label{W}
\eeq
where $j=u,\bar u, d,\bar d,\dots\ $.
The meanings of various factors in Eq.\,(\ref{W}) 
were discussed in detail
in \cite{nsy1999}. Here we should emphasize that at short distances,
{\it i.e.}, in the limit
$ b\ra 0 $, the function $\widetilde W_{ba}(b,Q,\widehat x, \widehat z, C_1,C_2,\mu_F)$ is
dominated by perturbatively calculable functions
${\cal C}^{in}_{ja}(\xh, C_1, C_2, b, \mu_F)$,
${\cal C}^{out}_{bj}(\zh, C_1, C_2, b, \mu_F)$, and the perturbative part
of the Sudakov factor $S^{P}(b,Q,C_1,C_2)$. The ${\cal O}(\alpha_s)$
expressions for these functions 
are listed in Appendix B.

In the region $b \gtrsim b_{max}$, where $b_{max}$ can be chosen to be 
$0.5 \mbox{ GeV}^{-1}$, 
nonperturbative contributions 
to $\widetilde W_{ba}$ become important. 
These nonperturbative contributions are absorbed into a function
$S^{NP}(b,Q,\dots)$, which vanishes in the limit $b\ra 0$ but dominates 
$\widetilde W_{ba}(b,Q,\widehat x, \widehat z, C_1,C_2,\mu_F)$ for $b \gtrsim b_{max}$.
Until a nonperturbative theory of strong interactions is developed, 
the explicit parametrization of $S^{NP}(b,Q,\dots)$ can only be
found from comparison with experimental data. One possible realization of
a smooth transition in the region $b\sim
b_{max}$ between the perturbative and nonperturbative parts
of $\widetilde W_{ba}(b,Q,\widehat x, \widehat z, C_1,C_2,\mu_F)$  
is to replace the parameter $b$ in 
the ${\cal C}$-functions and the perturbative Sudakov factor $S^{P}$
by the variable
\beq
\label{bstar}
b_* \equiv \frac{b}{\sqrt{1+ (b/b_{max})^2}},
\eeq
as written in (\ref{W}).

There is  a subtle difference between ${\cal C}^{in}$ 
and ${\cal C}^{out}$. While the initial-state 
coefficient functions  ${\cal C}^{in (1)}_{ba}(\xh,C_1,C_2,b,\mu_F)$ 
given in
Eqs.\,(\ref{C1in}) and (\ref{C1in2}) depend on the
factorization scale $\mu_F$ through a factor 
$\ln {[\mu_F b/b_0]}$, the final-state functions 
${\cal C}^{out (1)}_{ba}(\zh,C_1,C_2,b,\mu_F)$ 
given in Eqs.\,(\ref{C1out}) and (\ref{C1out2})  
depend instead on $\ln {[\mu_F b/(b_0 \zh)]}$. 
The additional term $\propto \ln{\zh}$ 
in the functions ${\cal C}^{out (1)}_{ba}(\zh,C_1,C_2,b,\mu_F)$
 becomes large and negative when $\zh \ra
0$, so that it can significantly influence the ${\cal
O}(\alpha_s)$  contribution at small values of $\zh$.
As a result, the resummed total rate tends 
to be lower than its fixed-order counterpart for $z \lesssim
0.1$. We will discuss this issue in more detail in Section\,\ref{SecChgd}.
Similarly, the ${\cal O}(\alpha_s)$ part of the NLO structure
function $\prescr{1}{\Vh^{NLO}_{ba,j}}$
in (\ref{V1NLO}) depends on $\mu_F$ through 
a logarithm  $\ln{[\mu_F^2/(\zh q_T^{sep})^2]}$.

The appearance of the additional terms $\propto \ln{\zh}$ in the 
functions ${\cal C}^{out (1)}_{bj}$ and  
$\prescr{1}{\Vh^{NLO}_{ba,j}}$
reflects the specifics of 
separation of the ${\cal O}(\alpha_s)$  ``hard''
cross section $(\widehat \sigma)_{hard}$ from the  collinear 
contributions to the FFs in the \MSbar factorization 
scheme. The easiest way to see the specific origin of the $\ln\zh$ terms
is to notice that the dependence
on the parameter $\mu$ in the $n$-dimensional expression
(\ref{NLL2}) for
$\prescr{1}\Vh_{ba}(\xh,\zh,Q^2,q_T^2)$
comes through a factor $(2\pi \mu/\zh)^{4-n}$, rather than through a
more conventional  $(2\pi \mu)^{4-n}$.
It is this factor that generates the
$\mu_F$-dependent logarithmic terms $\ln{[\mu_F b/(b_0 \zh)]}$ 
in the functions 
${\cal C}^{out (1)}_{bj}(\zh,C_1,C_2,b,\mu_F)$ 
and $\prescr{1}{\Vh^{NLO}_{ba,j}}$. 
The ${\cal C}^{in (1)}_{ja}$-functions 
do not include $\ln{\zh}$ because
they are evaluated
along the direction $ \zh=1 $ in the phase space.  
In contrast, nothing forbids such a term 
in the functions ${\cal C}^{out (1)}_{bj}$,
in which $\zh$ can be anything between $z$ and 1.
This contribution produces terms of the form $P_{bj} \ln \zh$ in 
${\cal C}^{out (1)}_{bj}(\zh,C_1,C_2,b,\mu_F)$,
which suppress the resummed rate for
the production of hadrons with very small energies. 

In Ref.\,\cite{nsy1999}, 
the terms proportional to $\ln{\widehat z}$ were left out of
the expressions for the ${\cal C}^{out}_{ba}$-functions. 
This error  
affected the value of the ${\cal C}^{out}$-function for the energy flow
in Ref.\,\cite{nsy1999}; the correct expression is Eq.\,(\ref{Coutz})
in this paper. We have found that 
theoretical distributions that  were shown in the Figures of
Ref.\,\cite{nsy1999} will be approximately reproduced with the correct
${\cal C}^{out}_{ba}$ 
if the following nonperturbative Sudakov factor is used: 
\beq
S^{NP}_z(b) = g_{BA}^{(2)}(b) \ln\frac{Q}{Q_0} 
+b^2 \biggl( -4.58 + \frac{0.58}{\sqrt{x}} \biggr).
\label{SNPzold}
\eeq
The explicit parametrization for
 the coefficient $g_{BA}^{(2)}(b)$ is given in \cite{nsy1999}. In the next
Section, we present an updated analysis of the transverse energy flow
that is compared to the newer data from the H1 Collaboration. We will
present a new parametrization of the nonperturbative Sudakov factor
which is in  better agreement with the whole set of experimental data.

\section{Resummation for the transverse energy flow}
\label{SecET}
As was discussed at length in \cite{nsy1999}, knowledge of the resummed 
semi-inclusive DIS cross section leads immediately to 
predictions for the pseudorapidity spectrum of the transverse energy
flow in the hadron Breit frame or the hCM frame. It is
advantageous to study the energy flows, because they are less dependent
on the specifics of final-state fragmentation of the scattered partons
into the observed hadrons. We therefore 
start the presentation of our numerical results with the comparison 
of the resummation formalism to the experimentally measured
pseudorapidity
distributions for the transverse energy flow in the hCM frame.


According to Eq.\,(\ref{qTeta}), 
the scale $q_T$ for massless particles is simply related to the
pseudorapidity  $\eta^{cm}$ in the hCM frame, if the energy of the
photon-proton collision is fixed. 
Therefore
the pseudorapidity distribution for the transverse energy flow
in the hCM frame $d\langle E_T\rangle/d\eta^{cm}$
can be converted into a $q_T$-distribution $d\langle E_T\rangle/d q_T$
by a simple change of variables.
From the viewpoint of theoretical analysis, it is preferable
to convert  $d\langle E_T\rangle/d\eta^{cm}$
into the $q_T$ distribution of the ``$z$-flow'' \cite{zflowdef}
that is defined as
\beq \frac{d
\Sigma_z}{d x \,dQ^2 \,dq_T} = \sum_B \int_{z_{min}}^1 \,z \,\frac{
d\sigma(e +A \ra e+ B+ X) }{dx \,dz \,dQ^2 \,dq_T}\,d z. 
\label{zflowdef}
\eeq
Under the approximation that all particles are massless, the distribution
$d\langle E_T\rangle /d\eta^{cm}$ is simply related to the distribution
$d\Sigma_z/dq_T$, by
\beq
\frac{d \langle E_T \rangle}{d\eta_{cm}} = q_T^2 \frac{d \Sigma_z}{d q_T}.
\label{Sz2ET}
\eeq

In \cite{nsy1999} the predictions of the resummation formalism were compared
to data on transverse energy flow $\langle E_T\rangle$ 
published in \cite{H1z1}.
In a recent publication \cite{H1z2},
the H1 Collaboration at the DESY-HERA $ep$
collider presented new data on the
distributions $d\langle E_T \rangle/d\eta^{cm}$ in the same bins of
$x$ and $Q^2$ as in \cite{H1z1} and, in addition, in several bins at
significantly higher $x$ and $Q^2$.
In this paper, we repeat the analysis
of \cite{nsy1999} to produce distributions $d\Sigma_z/dq_T$
for the bins of $x$ and $Q^2$ discussed in \cite{H1z2}.
The procedure of our numerical analysis is identical to the one
used in \cite{nsy1999}. Our calculations use the parameters of the HERA
$ep$ collider.  The energies of the proton and electron beams
are taken to be 820 GeV and 27.5 \mbox{GeV}, respectively.
The current numerical results are obtained  using  
CTEQ5M1 parton distribution functions \cite{CTEQ5}. 

By the factorization theorems of QCD, the  hadron-level $z$-flow
$\Sigma_z$ can be
written as the convolution of a parton-level $z$-flow $\widehat \Sigma_z$
with the PDFs, 
\beq
\frac{d\Sigma_z}{d x d Q^2 d q_T^2 d \varphi} = \sum_a \int_{x}^{1}
\frac{d\xi_a}{\xi_a} F_{a/A}(\xi_a, \mu_F) 
\frac{d\widehat \Sigma_z (\mu_F)}{d \xh d Q^2 d q_T^2 d \varphi}. 
\eeq
Similarly to the sDIS cross section, the $z$-flow can be expanded
in a sum over the leptonic angular functions $A_\rho (\psi, \varphi)$:
\beq
\frac{d \widehat \Sigma_z}{d \xh d Q^2 d q_T^2 d \varphi} = \sum_{\rho=1}^{4} 
\prescr{\rho} \Vh_{za}(\xh,Q^2,q_T^2)
A_{\rho}(\psi,\varphi), \label{angz} 
\eeq
where the structure functions $\prescr{\rho}\Vh_{za}(\xh,Q^2,q_T^2)$
for the $z$-flow are related to
the structure functions $\prescr{\rho}\Vh_{ba}(\xh,\zh,Q^2,q_T^2)$
for the sDIS cross section by 
\beq
\prescr{\rho}\Vh_{za}(\xh,Q^2,q_T^2) = \sum_b \int_{0}^1
\zh d\zh\, \prescr{\rho}\Vh_{ba}(\xh,\zh,Q^2,q_T^2).
\label{Vcsz}
\eeq

The resummed $z$-flow is calculated as
\begin{equation}
\label{resumz}
\frac{d \Sigma_z}{d x d Q^2 d q_T^2 d \varphi}=
\sFs \frac{ A_1 (\psi, \varphi)}{2}
\int \frac{d^2 b}{(2\pi)^2} e^{i\vec q_T \cdot \vec
b } \widetilde W_z (b, x, Q)+Y_z,
\end{equation}
where
\begin{equation}\label{Wz}
\widetilde W_z (b, x, Q)=\sum_j e_j^2
{\cal C}^{out}_z \,e^{-S_z (b, Q, x)}\,({\cal
C}_{ja}^{in} \circ F_{a/A}) (x, b_*, \mu_F).
\end{equation}
The functions ${\cal C}_{ja} ^{in}$ in (\ref{Wz}) are the same as in
(\ref{W}). The coefficient ${\cal C}^{out}_z$ is 
\beq
 {\cal C}^{out}_z = 1 +
\frac{\alpha_s}{\pi} C_F \Bigl( -\frac{7}{16} -\frac{\pi^2}{3}
-
\ln^2 \frac{e^{-3/4} C_1}{C_2 b_0}\Bigr),
\label{Coutz}
\eeq
where $b_0\equiv 2 e^{-\gamma_E}$ and $\gamma_E = 0.577\dots$ 
is the Euler constant. 
The parameter $b_*$, given by (\ref{bstar}) with $b_{max} = 0.5 \mbox{
 GeV}^{-1}$,
is introduced in (\ref{Wz}) 
to smoothly turn off the perturbative dynamics 
when $b$ exceeds $b_{max}$. The numerical results for
the resummed $z$-flow are obtained using  $C_1/C_2 =\mu_F b = b_0$.

\begin{figure}[p]
\epsfysize 20cm
\epsffile{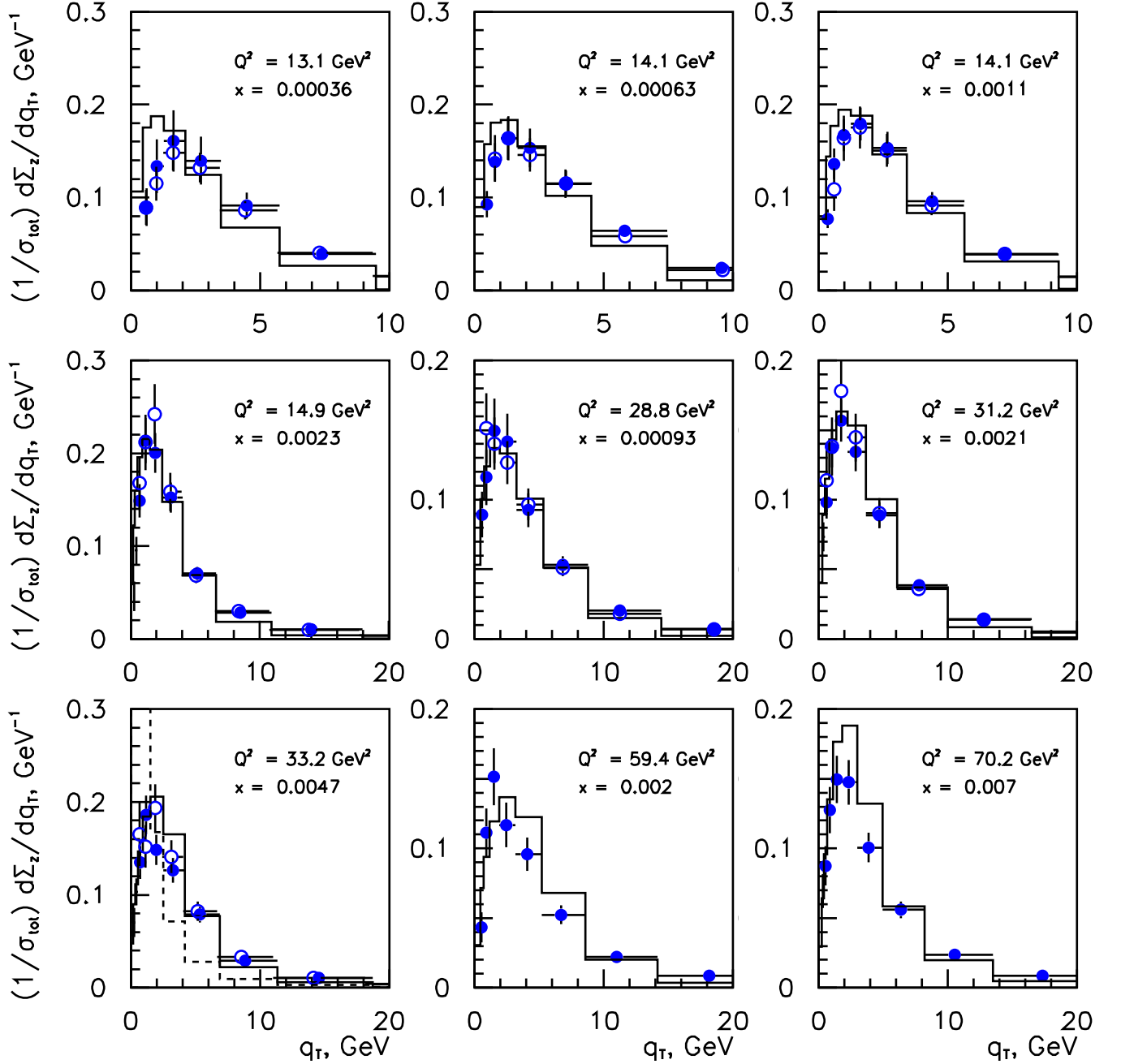}
\caption{Comparison of the resummed $z$-flow in the current region of
the
hCM frame 
with the data in the  low-$Q^2$ bins from 
Refs.\,\protect\cite{H1z2}
 (filled circles) 
and \protect\cite{H1z1}
(empty circles). The parametrization
of the nonperturbative Sudakov factor in (\ref{SNPz}) is used.
For the bin with $\langle Q^2\rangle=33.2
\mbox{ GeV}^2$ and  $\langle x \rangle = 0.0047$, the ${\cal O}(\alpha_s)$
contribution for $\mu_F=Q$ is shown as a dashed curve. 
\label{b17}
}
\end{figure}

\begin{figure}[p]
\epsfysize 20cm
\epsffile{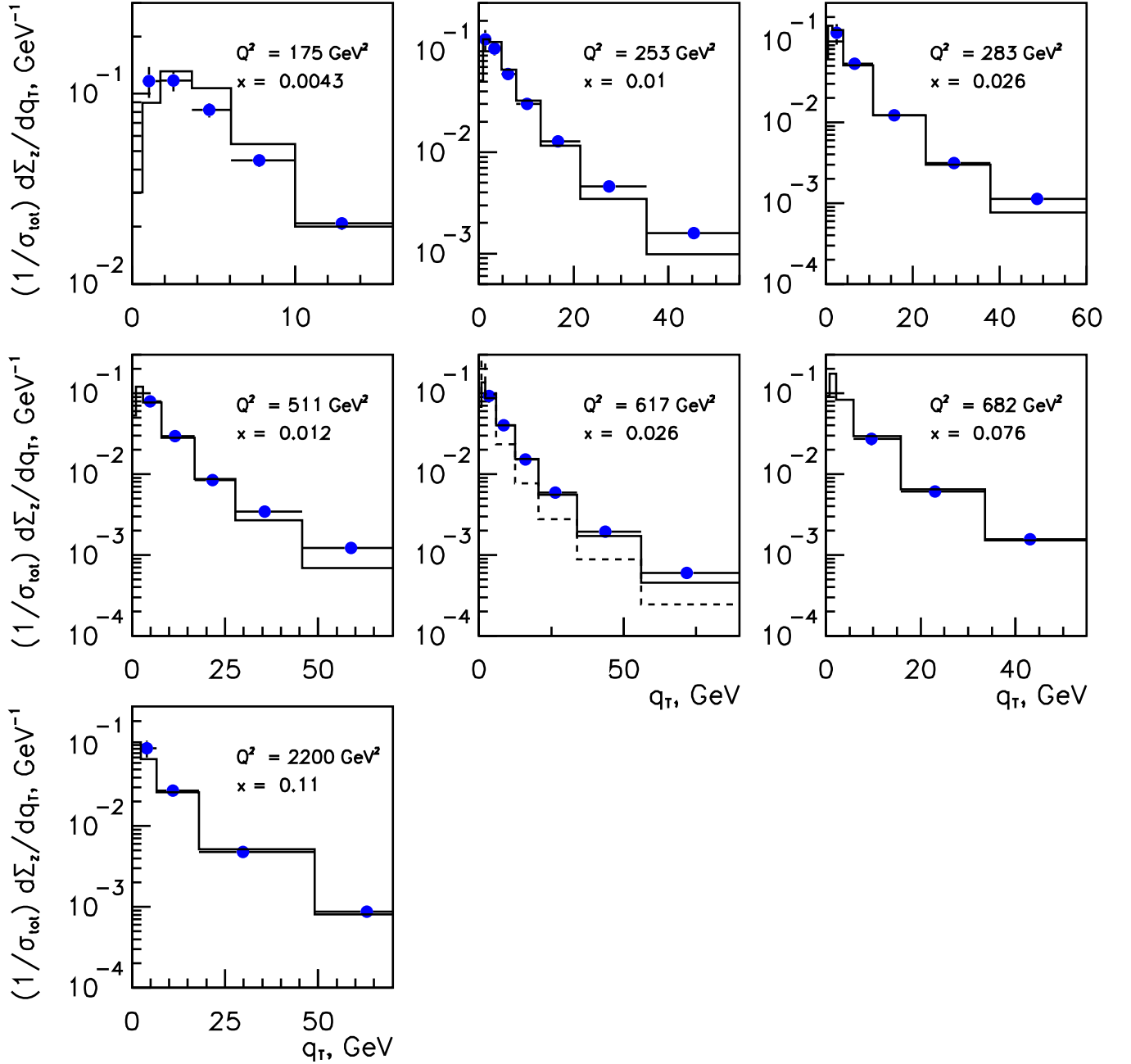}
\caption{Comparison of the resummed $z$-flow in the current region of 
the hCM frame with the data in the
high-$Q^2$ bins from Ref.\,\protect\cite{H1z2}. The parametrization
of the nonperturbative Sudakov factor in (\ref{SNPz}) is used.
For the bin with $\langle Q^2\rangle=617
\mbox{ GeV}^2$ and $\langle x \rangle = 0.026$, the ${\cal O}(\alpha_s)$
contribution for $\mu_F=Q$ is shown as a dashed curve. 
\label{b47}
}
\end{figure}

The term $Y_z$ in (\ref{resumz}) 
is the difference between the complete fixed-order expression at
${\cal O}(\alpha_s)$ for 
$d \Sigma_z/(d x d Q^2 d q_T^2 d \varphi)$ and its most
singular part 
in the limit $q_T \ra 0$;
that is,
\beq
Y_z = \frac{d \Sigma_z}{d x d Q^2 d q_T^2 d \varphi} - 
\Biggl( \frac{d \Sigma_z}{ d x d Q^2 d q_T^2 d\varphi}
\Biggr)_{sing}.
\eeq
The singular part calculated to ${\cal
O}(\alpha_s)$ is
\bea
\Biggl(\frac{d \Sigma_z}{ d x d Q^2 d q_T^2 d\varphi}
\Biggr)_{sing}&=& \sFs \alpi \frac{1}{2 q_T^2}  \frac{A_1(\psi, \varphi)}{2
\pi}
\nonumber \\
&\times& \sum_{j}
e_j^2 \Biggl[ \Bigl\{ (P_{qq} \circ F_{j/A})(x,\mu_F)
+(P_{qg}\circ F_{g/A})(x,\mu_F)\Bigr\}
\nonumber \\
&+& 2 F_{j/A}(x,\mu_F)  \Bigl\{C_F \ln
\frac{Q^2}{q_T^2}-\frac{3}{2}C_F\Bigr\} +{\cal
  O}\Bigl(\biggl(\frac{\alpha_s}{\pi}\biggr)^2, q_T^2\Bigr)\Biggr].
\label{asymz}
\eea
The formula (\ref{resumz}) is expected to provide a smooth matching
between the resummed $z$-flow in the region $q_T \ll Q$ and the
fixed-order $z$-flow in the region $q_T \gtrsim Q$. However,
we find that the ${\cal O}(\alpha_s)$ fixed-order $z$-flow
significantly undershoots the data in the region $q_T/Q \gtrsim 1$.
The resummed $z$-flow shows better agreement with the data, over a
wide range of $q_T/Q$, but also lies below the data if $q_T/Q$
significantly exceeds unity.
The discrepancy between the theoretical $z$-flows and the
data at $q_T/Q \gtrsim 1$ is illustrated for the bin $\langle
Q^2\rangle =33.2 
\mbox{ GeV}^2,\,\langle x \rangle = 0.0047$ in Fig.\,\ref{b17} and for
the bin \quad $\langle Q^2\rangle =617
\mbox{ GeV}^2,\,\langle x \rangle = 0.026$ in Fig.\,\ref{b47}.  
The data at sufficiently high $q_T$ lies above
the ${\cal O}(\alpha_s)$ $z$-flow and the resummed $z$-flow
in both bins.
This difference signals that 
calculation of higher order radiative corrections is required to achieve
agreement between the theory and the data for $q_T/Q \gtrsim 1$. 
The better consistency between the resummed $z$-flow and the data 
suggests that we should use the resummed $z$-flow up to values of
$q_T/Q\sim 1$, {\it i.e.,} without switching to the fixed-order 
expression. We have followed this procedure 
in the derivation of our numerical results.

We have also found that the agreement
between the resummed $z$-flow and the data 
in the region $q_T/Q \sim 1$ can be
improved by making an additional correction that accounts for the reduction
of the allowed phase space with the increase of $q_T$. This correction
was discussed in detail in \cite{nsy1999}. Its importance can be
understood by noting the steep rise of the PDFs as $x$ decreases,
at values of $x$  relevant to the HERA measurements. 
This steep rise in the small $x$ region, 
together with a mismatch between the phase
space in the $b$-space integral of the resummed
$z$-flow (\ref{resumz}) and that in the exact fixed-order
formula, may produce large numerical effects if $q_T$ is of
the order $Q$ or larger. 
To correct for this mismatch between the resummation and the perturbative
phase spaces, $x =
Q^2/(2 p_A\cdot q)$ is replaced in 
Eqs.\,(\ref{Wz}) and (\ref{asymz}) by the $q_T$-dependent quantity
\beq
\widetilde x = \frac{Q^2 + q_T^2}{Q^2} x.
\label{kinem}
\eeq
The replacement $x\ra \widetilde x$
preserves the leading-order kinematics when
$q_T$ approaches zero; at the same time, 
it simulates the reduction of the allowed
phase space when $q_T$
becomes comparable with $Q$. It affects only the convolution integrals 
$(P_{ba} \circ F_{a/A})$ and $({\cal C}_{ba} \circ F_{a/A})$, 
in which the $q_T$-independent lower limit of integration $x$ is
replaced by the variable $\widetilde x$. The result of this 
replacement is that the convolution
integrals at large $q_T$ do not include large contributions from very
small values of $x$, which helps to improve the accuracy of 
the matching procedure at $q_T \sim Q$.

Before turning to the discussion  of the experimental data, let us
comment on the structure of the Sudakov factor $S_z(b,Q,x)$ in
(\ref{Wz}). As in the case of the sDIS cross section, the
Sudakov factor for the $z$-flow consists of both perturbative and
nonperturbative parts,
\bea
S_z (b, Q, x) &=& \int_{C_{1}^{2}/b^{2}_*}^{C_{2}^{2}Q^{2}}
\frac{d\ov \mu^{2}}{\ov \mu^{2}}\Biggl(A(\alpha_s(\ov \mu), C_1)
\ln\frac{C_{2}^{2}Q^{2}}{\ov \mu^{2}} + B (\alpha_s(\ov \mu ),
C_1,C_2)\Biggr) \nonumber\\
&+& S_{z}^{NP}(b,Q,x).
\eea 
A realistic parametrization for the 
nonperturbative Sudakov factor $S_{z}^{NP}(b,Q,x)$ can be obtained
by comparison with  experimental data, especially with the measured
pseudorapidity distributions at $Q \approx 3-20$ GeV. As $Q$
increases, 
we expect the data to become less sensitive to 
nonperturbative effects and to be dominated by
perturbatively calculable parton radiation. In
\cite{nsy1999} we found a parametrization of $S_{z}^{NP}(b,Q,x)$ 
that was consistent with the data published in \cite{H1z1}. But, 
because of the limited scope of those data, that parametrization was
expected to be valid only for $10^{-4} \leq x\leq 10^{-2}$. 
Also, the $Q$ dependence of   $S_{z}^{NP}(b,Q,x)$ could not be tested,
since all the pseudorapidity distributions in Ref.\,\cite{H1z1} were presented
in a small range of $Q \approx 2 -6$ GeV. Hence, 
the parametrization of $S_{z}^{NP}(b,Q,x)$ in \cite{nsy1999} used a
model prediction for the coefficient of $\ln Q/Q_0$ based on
the assumption of crossing symmetry between 
the nonperturbative Sudakov factors in sDIS, $e^+ e^-$ hadroproduction,
and Drell-Yan processes.

In its more recent publication \cite{H1z2}, the H1 Collaboration 
presented pseudorapidity distributions of the transverse
energy flow for $Q^2$ up to  $2200 \mbox{ GeV}^2$. 
However, as just mentioned, the data at such
high $Q^2$ should be rather insensitive to the nonperturbative
dynamics, {\it e.g.,} because the nonperturbative effects would be
important only at very large values of $\eta^{cm}$, 
for which the resolution of the detector is poor. Thus the data at
very high $Q^2$ is not very informative about the $Q^2$-dependence of
$S_{z}^{NP}(b,Q,x)$ either. To learn about the dependence on $Q^2$, 
one needs to look at data in the range $Q \sim 5-20$ GeV.
Fortunately, the H1 Collaboration presented  distributions in two bins at
intermediate values of $\langle Q^2\rangle$, namely
$\langle Q^2 \rangle =  59.4 \mbox{ GeV}^2$ and 
$\langle Q^2\rangle= 70.2 \mbox{ GeV}^2$.
Together with the data from Refs.\,\cite{H1z1,H1z2} at lower values of
$Q$, these distributions provide the first direct tests of  
the $Q^2$-dependence of $S_{z}^{NP}(b,Q,x)$.
We therefore present a numerical value for the coefficient of $\ln Q/Q_0$
that yields reasonable agreement with all of the analyzed data.
The resulting value of this 
coefficient differs significantly from the model parametrization in 
\cite{nsy1999}. However, we should not draw too strong a conclusion from
this difference, because it
might be caused by ambiguities 
in the separation of $Q^2$ dependence 
and $x$ dependence in the existing data. To draw a strong
conclusion
about the crossing symmetry model, experimental 
pseudorapidity distributions in a larger range of $x$ at intermediate
values of  $Q^2$ will be needed.

From our analysis, we find that the data from Refs.\,\cite{H1z1,H1z2} are
consistent with the following parametrization of the nonperturbative
Sudakov factor:
\beq
S^{NP}_{z}(b, Q, x) = 
b^2\,\left (0.013 \frac{(1-x)^3}{x} +0.19\ln{\frac{Q}{Q_0}}+ C\right),
\label{SNPz}
\eeq
where the parameter $Q_0$ is fixed to be 2 GeV, and where we set $C=0$
for reasons explained later.
In Figs.\,\ref{b17} and \ref{b47} we compare the resummed $z$-flow,
calculated according to (\ref{resumz}) and using the parametrization 
$S^{NP}_z$ in (\ref{SNPz}), to the data from \cite{H1z2}. 
The $z$-flow is normalized by the inclusive DIS cross section 
$d\sigma/(d x d Q^2)$ (which will be denoted by the brief notation 
$\sigma_{tot}$).
Following
the recent H1 publication \cite{H1z2}, we organize these data into
two sets of bins for fixed $x$ and $Q^2$. One set covers 
$3.2 <\langle Q^2 \rangle <
70.2 \mbox{ GeV}^2$ and 
$8\times 10^{-5} < \langle x\rangle < 7\times 10^{-3}$, out of which
we consider the bins with $\langle Q^2 \rangle \geq 13.1 \mbox{ GeV}^2$. 
The other set covers  
$175 <\langle Q^2\rangle <
2200 \mbox{ GeV}^2$ and $0.0043 < \langle x\rangle < 0.11$.  
The transformation of the experimental 
$d\langle E_T \rangle/d\eta^{cm}$ distributions into 
$d\Sigma_z/dq_T$ distributions 
and the calculation of the theoretical distributions 
$d\Sigma_z/dq_T$ are done using the
mean values $\langle x\rangle$ and 
$\langle Q^2\rangle$ for each bin published in \cite{H1z2}.
In seven of the nine low-$Q^2$ bins (Fig.\,\ref{b17}) 
we also show the earlier published data from \cite{H1z1}. 

Let us discuss the features of the data presented in
Figs.\,\ref{b17} and \ref{b47}. First, the data in the low-$Q^2$ bins is
significantly influenced by nonperturbative effects and
therefore is sensitive to the details of the parametrization of the 
nonperturbative Sudakov factor $S^{NP}_z(b,Q,x)$. This feature
can be seen from the
abundance of data points around the maximum of the $q_T$-distribution,
where the shape is mainly 
determined by the nonperturbative Sudakov factor.
Also, the low-$Q^2$ data
from HERA is characterized by small values of $x$, between $10^{-4}$
and $10^{-2}$. 
As we noticed in our earlier work
\cite{nsy1999}, for the theory to be consistent with the data in this range
of $x$, the nonperturbative Sudakov factor must increase rapidly as 
$x \ra 0$, at least as $1/\sqrt{x}$ ({\it cf.} Eq.\,(\ref{SNPzold})).  
In our new analysis,
we find that growth of $S^{NP}_z(b,Q,x)$
as $1/x$ at small $x$ 
is in better agreement with the more recent data from \cite{H1z2}.

Secondly, the data in the high-$Q^2$ bins of Fig.\,\ref{b47}
shows a behavior that is qualitatively different from
Fig.\,\ref{b17}. 
In the region 
covered by the experimental data points, 
the $q_T$ distribution 
is a monotonically decreasing function of $q_T$, which
shows good agreement with the resummed 
$z$-flow over a significant 
range\footnote{We point out once again that 
both the ${\cal O}(\alpha_s)$ and resummed $z$-flow lie below the data
at very large $q_T$, in all bins of $x$ and $Q^2$ in Figures\,\ref{b17} and
\ref{b47}.}
of $q_T$. In the region $q_T < 10$ GeV, {\it i.e.}, where the 
maximum of the $q_T$ distribution is located and where
nonperturbative effects are important, the experimental
$q_T$-bins are too large to provide any information about the shape of 
$d\Sigma_z/d q_T$.    Thus, as mentioned earlier, 
the published high-$Q^2$ $z$-flow data from Ref.\,\cite{H1z2}
is not sensitive to the
dynamics described by the nonperturbative
Sudakov factor $S^{NP}_z(b,Q,x)$.  

A third comment is that most of 
the high-$Q^2$ data points in Fig.\,\ref{b47} 
correspond to $\langle x \rangle > 10^{-2}$. If the resolution of the H1
measurements at high $Q^2$ were better in the small-$q_T$ region, 
then the high-$Q^2$ data would also reveal  
the behavior of $S^{NP}_z (b,Q,x)$ at large $x$. 
But, as mentioned before, the published data in the high-$Q^2$ bins
are not very sensitive to the shape of the $z$-flow 
at small $q_T$. Therefore it is not possible to impose any
constraints on $S^{NP}_z (b,Q,x)$ at large values of $x$,
except that it should be positive, $S^{NP}_z (b,Q,x) > 0$. For this
reason we
have chosen a parametrization of the $x$-dependent part 
of $S^{NP}_z (b,Q,x)$ in (\ref{SNPz}) such that $S^{NP}_z (b,Q,x)$ grows
approximately as $1/x$ as $x \ra 0$ and is positive for all $x$.
For the same reason, we chose $C=0$ in  
the parametrization (\ref{SNPz}). Although the most general parametrization
of $S_z^{NP}(b,Q,x)$ can have $C \neq 0$, the current data 
cannot distinguish 
between the parametrization (\ref{SNPz}) with $C=0$ and $C\neq 0$, 
as long as the value of $C$ is not very large.

\subsection*{How trustworthy is the resummed $z$-flow at large $q_T$?}
As noted earlier,
the ${\cal O}(\alpha_s)$ fixed-order $z$-flow is much
larger than the data in the region $q_T/Q \ll 1$ and
much smaller than the data in the region $q_T/Q
\gtrsim 1$. In the small-$q_T$
region, the resummed $z$-flow is, by its construction,
more reliable than the fixed-order result.  In the large-$q_T$ region,
the resummed $z$-flow, with 
the kinematical correction (\ref{kinem}) included,
 is also in  better agreement with the
data than the fixed-order calculation. But theoretically, the resummed 
$z$-flow at large $q_T/Q$ is not absolutely trustworthy,
because it does not include those parts of the fixed-order $z$-flow
that are subleading in the limit $q_T \ra 0$, but which might be important
at large $q_T$. If the NLO result were in a good
agreement with the data at large $q_T$, it would be justified 
to consider it a more reliable prediction in this region. But since
the ${\cal O}(\alpha_s)$ contribution is systematically smaller
than the data,
higher-order corrections are presumably necessary in order to
describe the region $q_T \gtrsim Q$ reliably.

A systematic approach for improving the
theoretical description of the large-$q_T$ region would require inclusion
of the complete
${\cal O}(\alpha_s^2)$ terms in both the fixed-order and
resummed $z$-flows. But because such a 
calculation is not available, it might be beneficial to use 
the resummed $z$-flow as a better theoretical prediction both in
the region $q_T/Q \ll 1$, where application of the resummation formalism
is fully justified, and for $q_T$ up to several units of $Q$, where the
resummed $z$-flow agrees with the data better than the
fixed-order one.  Then the use of the resummed $q_T$-distributions
of the $z$-flow will provide more reliable predictions for other
observables relevant to the sDIS process.

As an example, resummation can improve the reliability of the
theoretical prediction for the azimuthal asymmetry of the $z$-flow. 
The $b$-space resummation formalism 
affects only the coefficient 
$\prescr{1}V_{zA}$ of the
angular function $A_1 (\psi, \varphi)$. 
This coefficient is the one that dominates
the $\varphi$-integrated 
$z$-flow in the small-$q_T$ region, where the energy flow is the
most intense. 
On the other hand, the main goal of the measurement of angular
asymmetries is to study structure functions other than $\prescr{1}V_{zA}$,
{\it e.g.,} those corresponding to the angular functions 
$A_3(\psi, \varphi) =  - \cos \varphi \sinh 2\psi $ and 
$A_4 (\psi, \varphi) = \cos 2 \varphi \sinh^2 \psi $. 
By using a better approximation for the coefficient $\prescr{1}{V}_{zA}$, 
it is possible to measure the coefficients $\prescr{3,4}V_{zA}$ more
reliably. 
Conversely, by knowing that the all-order
resummation effects are important in the region of
small $q_T$ and by concentrating on the
region where $q_T$ is of the order $Q$ or larger, one may find 
angular asymmetries that are well approximated in 
the lowest orders of perturbative QCD. We discuss 
the impact of resummation on the angular asymmetries  
in more detail in another paper \cite{nsyasym}.

\section{Distributions of charged particle multiplicity}
\label{SecChgd}
Let us now turn to the discussion of  semi-inclusive DIS cross sections.
Although the resummation formalism, as outlined in
Section\,\ref{SecSummary},  can describe the cross section for
any massless final-state particle (provided that the fragmentation
functions for this particle are known), in this
Section we will concentrate on distributions of the 
charged particle multiplicity, defined as
\beq
\frac{1}{\sigma_{tot}} 
\frac{d\sigma (A + l \ra h^{\pm} + l^{\prime} + X)}{d\Theta}.
\eeq
Here
$\Theta $ is some kinematical variable,
such as the variable  $q_T^2$ in Eq.\,(\ref{qT2}),  
the transverse momentum $p_T$ 
of the final-state charged particle in the hCM frame,
or the Feynman variable $x_F$, defined by   
\beq
x_F \equiv \frac{2 p_{B}^z}{W} = z  \left(1-\frac{q_T^2}{W^2}\right).
\label{xF}
\eeq
In (\ref{xF}) 
$p_{B}^z$ is the longitudinal component of the momentum of the
final-state 
hadron in the hCM frame. 
For small values of $q_T$, {\it i.e.,} in the region with the highest rate,
\beq
x_F \approx z.
\eeq

In our calculation we assume that the charged
particles registered in the detector are
mostly charged pions, kaons and protons. Therefore
the cross section for charged particle production
can be calculated using (\ref{factcs}) with the replacement
of the fragmentation functions 
$D_{B/b}(\xi_b,\mu_F)$ 
by
\beq
D_{h^{\pm}/b}(\xi_b,\mu_F) = \sum_{B} 
D_{B/b}(\xi_b,\mu_F). 
\eeq
The fragmentation functions $D_{B/b}(\xi_b,\mu_F)$ are
known reasonably well only for $\xi_b
\gtrsim 0.05 - 0.1$ \cite{BKK,Kretzer,Bourhis}. 
Thus, the formalism presented here
is applicable to the production of charged particles with sufficiently
large energies, {\it i.e.}, for $z \gtrsim 0.05$. 

Certain experimental distributions 
are readily available from the literature \cite{ZEUSchgd96,EMC,E665},
such as
$d\sigma/dp_T$, $d\sigma/d x_F$, 
as well as 
distributions for the average transverse momentum $\langle p_T^2
\rangle$. However, 
the ``experimental'' $q_T$ distributions must currently be 
derived from pseudorapidity distributions by using Eq.\,(\ref{qTeta}).
Although the distributions $d\sigma/dp_T$ and $ \langle p_T^2
\rangle$ are quite sensitive to resummation effects, 
they cannot be interpreted as easily as the distributions $d\sigma/dq_T$, 
primarily because the  distributions
$d\sigma/dp_T$ and $ \langle p_T^2 \rangle$  mix  
resummation effects at small values of $q_T$ 
with perturbative contributions from the region $q_T/Q \gtrsim 1$. 
The most straightforward way to study the effects of
multiple parton radiation would be to consider
the $q_T$ (or pseudorapidity) distributions 
that satisfy the additional
requirement $z >0.05-0.1$ and that 
are organized into small bins of $Q^2$ and $x$.
Unfortunately, such distributions
have not been published yet. 

Because the experimental $q_T$ distributions are unavailable, 
we have decided to undertake a simpler analysis than the one presented for 
the energy flow. Our  goal here is to understand how 
the multiple parton radiation {\it could} affect various aspects of charged
particle production.  
For this purpose we focus our attention on data from the ZEUS
Collaboration \cite{ZEUSchgd96},
which presents the charged particle multiplicity in a phase-space
region characterized by the mean values  $\langle
W \rangle = 120 \mbox{ GeV} $, $\langle Q^2 \rangle = 28 \mbox{ GeV}^2$,
and the additional constraint $z > 0.05$. These values of $\langle W
\rangle$ and $\langle Q^2 \rangle$ 
translate into an average value of $x= 1.94\times 10^{-3}$. 
Within the framework of a simple model for nonperturbative effects at
small $q_T$, we will demonstrate that resummation 
describes qualitative features of this set of experimental data
better than the fixed-order calculation.
In all of the cases presented, our
strategy is to compare the resummed multiplicity to that from the
next-to-leading order calculation. 
In the numerical analysis, the
multiplicity was calculated using the CTEQ5M1 PDFs \cite{CTEQ5}
and the FFs from \cite{Kretzer}. For the resummed multiplicity, the
``canonical'' combination $C_1/C_2 =\mu_F b = 2
 e^{-\gamma_E}$ was used. Unless specified,
the NLO cross section was calculated according to (\ref{pert}),
with the parameter $q_T^{sep} = 1.2$ GeV, 
and for the factorization scale $\mu_F = Q$.

As in the case of the $z$-flow, the resummed charged particle multiplicity may
suffer from matching ambiguities at $q_T/Q \sim 1$. 
In Section\,\ref{SecET}, we found that the resummed $z$-flow is in 
better agreement with the experimental $d\sigma/dq_T$ distributions
than the NLO $z$-flow, for the whole range 
$q_T/Q \lesssim 2-4$. That result suggests
that it might be preferable to
use the resummed $z$-flow in the whole range $q_T/Q \lesssim 2-4$
as a better theoretical prediction, until the ${\cal O}(\alpha_s^2)$
prediction for the $z$-flow in the region $q_T/Q \gtrsim 1$ becomes 
available.
In the case of the charged particle multiplicity, 
the resummed cross section, which is 
calculated according to the formula
\begin{equation}
\frac{d \sigma_{BA}}{d x d z d Q^2 d q_T^2 d \varphi}=
\sFs \frac{ A_1 (\psi, \varphi)}{2}
\int \frac{d^{2}b}{(2\pi)^{2}} e^{i\vec q_T \cdot \vec
b } \widetilde W_{BA}(b, x, z , Q)+Y_{BA},
\end{equation}
overestimates the experimentally measured rate for the production of
charged particles with $p_T > 2$ GeV. 
This discrepancy
indicates that the resummed cross section in the region $q_T/Q \gtrsim
1$ is too high, so that switching to the perturbative cross section in
this region is in fact required. 
Therefore, we have chosen to use the resummed cross section for 
$q_T \leq 5$ GeV
and switch to the next-to-leading order cross section for $q_T \geq 5$
GeV. 

As in the case of the $z$-flow, the shape of the $q_T$ distribution for
the charged particle multiplicity at small values of $q_T$ 
depends strongly on the unknown nonperturbative Sudakov factor $S^{NP}(b, Q,
x, z)$. 
For the purposes of this study, we 
introduce a preliminary parametrization of the nonperturbative
Sudakov factor for the {\it fixed} values of $x = 1.94 \times 10^{-3}$
and $Q^2 = 28 \mbox{ GeV}^2$, {\it i.e.,} the values that coincide with the
average values of $x$ and $Q^2$ in \cite{ZEUSchgd96}. 
This {\it $z$-dependent} 
parametrization is 
\beq
S^{NP} \left (b, 
Q^2  = 28\mbox{ GeV}^2, 
x =1.94\times 10^{-3},
z\right) = 
b^2 \left( 0.18 + 0.8 \frac{(1-z)^3}{z^{1.4}}\right).
\label{SNPchgd}
\eeq 
Since the ZEUS Collaboration did not publish pseudorapidity
distributions for the charged particle multiplicity $(1/\sigma_{tot}) d
\sigma/d\eta^{cm}$ in bins of varying $z$, we had to deduce
information about the $z$-dependence of 
$S^{NP}$ from the less direct
distribution of $\langle p_T^2 \rangle $ vs.~$x_F$ presented in Fig.\,3c
of \cite{ZEUSchgd96}. This distribution,  
known as a ``seagull'' for its
characteristic shape (Fig.\,\ref{seagull}a), 
can be converted into the more illustrative
distribution of $\langle q_T^2 \rangle$ vs.~$x_F$ (Fig.\,\ref{seagull}b).
Since the major portion of the registered events comes from the region
$q_T^2/W^2 \ll 1$, or $x_F \approx z$, a first estimate of 
the experimental data points  for the distribution of 
$\langle q_T^2 \rangle$ vs.~$x_F$ can be obtained by assuming that
\beq
\langle q_T^2 \rangle \approx \frac{\langle p_T^2 \rangle}{\langle
z\rangle^2} \approx
\frac{\langle p_T^2 \rangle}{\langle
x_F\rangle^2},
\label{pT2toqT2}
\eeq
where $\langle x_F \rangle $ denotes central values of $x_F$ in each
bin in Fig.\,\ref{seagull}a.\footnote{ In principle, 
a more accurate experimental distribution $\langle q_T^2
\rangle$ vs.~$x_F$ can be determined by its direct measurement.} We
refer to the resulting values as ``derived data''.

Note that the shapes of $\langle p_T^2 \rangle$ vs. $x_F$ 
and $\langle q_T^2 \rangle$ vs. $x_F$ are quite different.
The transformation from Fig.\,\ref{seagull}a to Fig.\,\ref{seagull}b
shows immediately that the wing-like shape of the
distribution of $\langle p_T^2 \rangle$~vs.~$x_F$ should be attributed to
a purely kinematical effect, namely an extra factor $1/z^2$ which is absent
in the distribution of $\langle q_T^2 \rangle$ vs.~$x_F$. Once this extra
factor is removed, we see from Fig.\,\ref{seagull}b that 
$\langle q_T^2 \rangle$ increases monotonically and rapidly as $z$ approaches
zero. In other words, the $q_T$ distribution
broadens rapidly when $z$ decreases. This behavior is
approximately realized by the simple 
$z$-dependent nonperturbative Sudakov
factor $S^{NP}(b,Q,x,z)$  given in 
Eq.\,(\ref{SNPchgd}).

The shape of the resummed cross section at $q_T \leq 5$ GeV and the
parametrization of $S^{NP}(b,Q,x,z)$ were chosen to maximize the agreement
between the resummed distribution of 
$\langle q_T^2 \rangle$~vs.~$x_F$ 
and the ``derived data'' ({\it cf.} Fig.\,\ref{seagull}b).
Figure\,\ref{seagull}b shows that the resummed calculation is in  
better agreement with
the data points than the NLO expression. 
We have found it difficult to 
reproduce the rapid growth of $\langle q_T^2 \rangle$ as $x_F \ra 0$
in either approach. In the future, it will be interesting 
to see how a more precise theoretical study will be able to adequately 
explain this rapid growth of $\langle q_T^2 \rangle$ in the region 
$x_F \ra 0$, assuming that the actual experimental data  
for the $\langle q_T^2 \rangle$~vs.~$x_F$
distribution resemble the
``derived data'' discussed above.

\begin{figure}[htb]
\hspace{3cm}
\epsfysize 12cm
\epsffile{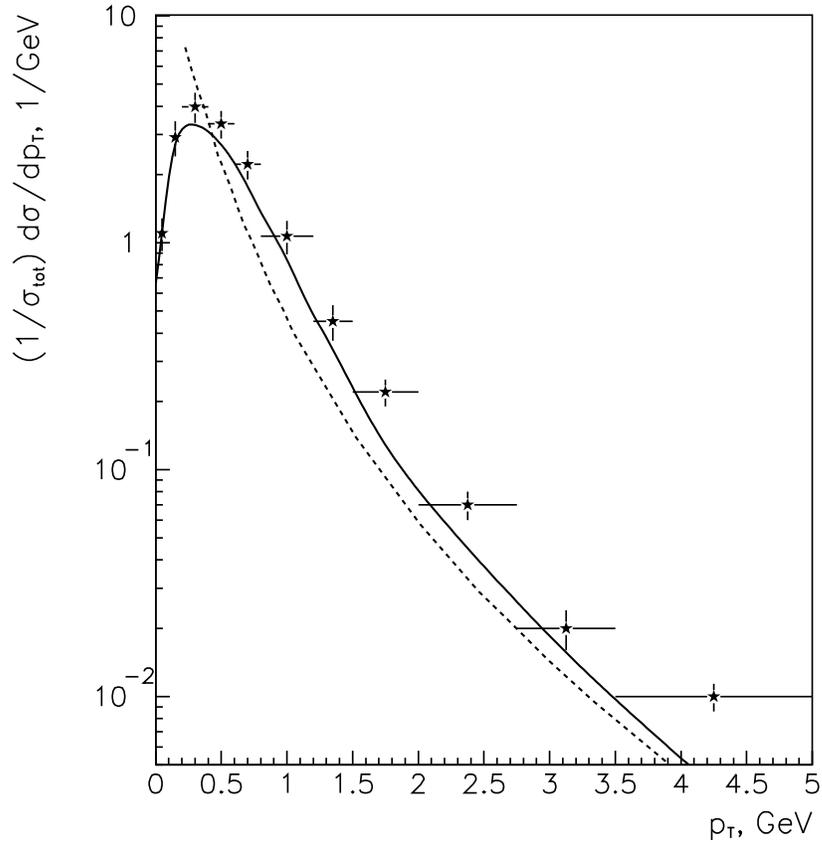}
\caption{
\label{dsigdpt}
The dependence of the charged particle multiplicity on the
transverse momentum $p_T$ of the observed particles in the hCM frame.
The data points are from \protect\cite{ZEUSchgd96}. The solid and
dashed curves correspond to the resummed and NLO multiplicities,
respectively.} 
\end{figure}

The resummation also significantly  affects the 
$p_T$ dependence of the charged particle multiplicity. In
Fig.\,\ref{dsigdpt} we present the distribution $(1/\sigma_{tot})
d\sigma/dp_T$. We see that resummation effects must be
included to describe the shape of this distribution at $p_T \leq
1$ GeV. Furthermore, resummation also improves the
agreement between the theory and the experiment in the whole range of
$p_T$. 
Through Eq.\,(\ref{pTB}),
the improved description of the $q_T$ distribution in the
small-$q_T$ region translates into a better agreement with the $p_T$
distribution in the whole range of $p_T$. Just as in the case of
the $z$-flow, the fixed-order calculation gives a rate that is too
small compared to the data, which
implies that higher-order corrections are important. 
Until the complete ${\cal O}(\alpha_s^2)$ corrections are available, 
the resummation formalism, which already
accounts for the most important contributions in the region of the phase space
with the highest rate ({\it i.e.,} at small $q_T$), 
serves as a better theoretical prediction  
in the whole range of $p_T$.

\begin{figure}[htb]
\hspace{2cm}
\epsfysize 12cm
\epsffile{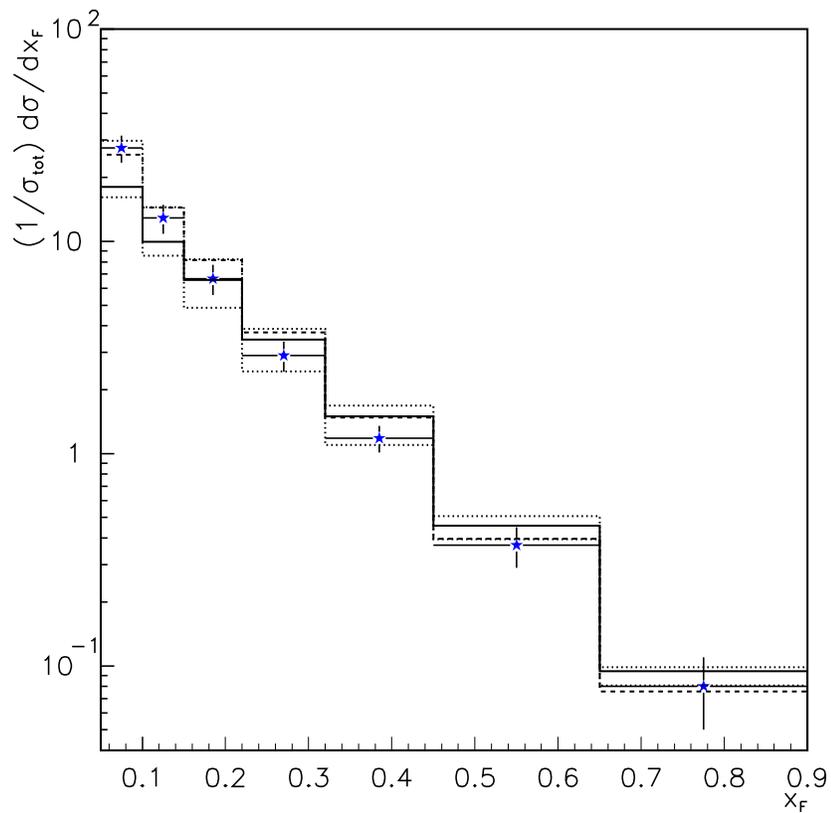}
\caption{\label{dsigdxf}
The dependence of the charged particle multiplicity on the
Feynman variable $x_F$ in the hCM frame. The solid curve corresponds
to the resummed multiplicity. The dashed, lower dotted
and upper dotted curves correspond to the NLO multiplicity calculated
for $\mu_F = Q,\ 0.5 Q$ and $2 Q$, respectively.} 
\end{figure}

Finally, Fig.\,\ref{dsigdxf} shows the $x_F$-distribution for the
charged particle multiplicity $(1/\sigma_{tot})d\sigma/dx_F$. 
We see that both the
resummed and fixed-order distributions are in reasonable
agreement with the data and with earlier published theoretical
results for the ${\cal O}(\alpha_s)$ $x_F$-distributions \cite{Graudenz}. 
For the fixed-order multiplicity, we present two additional curves
corresponding to  different choices of the factorization scale $\mu_F$ in
(\ref{factcs}); the lower and upper dotted curves correspond to $\mu_F
=0.5 Q$ and $2 Q$, respectively.
Note that the scale dependence of the NLO
multiplicity increases when
$z\ra 0$. Also note that the resummed multiplicity is significantly
lower than the data in the two lowest bins of $x_F$ ($\langle x_F \rangle =
0.075$ and $0.125$), but consistent with the NLO
multiplicity within the uncertainty due to the scale dependence.   
Such behavior of the resummed multiplicity  
results from the dependence of the ${\cal O}(\alpha_s)$ 
coefficient functions
${\cal C}^{out (1)}_{ba}(\widehat z,C_1,C_2,b_*,\mu_F)$ 
on the additional term
$\ln\zh$ which was given in Eqs.\,(\ref{C1out}) and (\ref{C1out2}) and 
discussed at the end of Section\,\ref{SecSummary}. 
This negative logarithm dominates the ${\cal C}^{out (1)}$-functions
at very small values of $\zh$. Similarly,
the integral (\ref{NLOcs}) of the NLO cross section 
over the lowest bin $0 \leq q_T^2 \leq (q_T^{sep})^2 $ 
depends on $\ln{\zh}$ through the terms
\bdm
    -\frac{\alpha_s}{2\pi}\ln{\frac{\mu_F^2}{(\zh \,q_T^{sep})^2}}\biggl(
       \delta(1-\zh) \delta_{b j} P_{j a} (\xh)  
      + P_{b j} (\zh)\delta(1-\xh)\delta_{j a} 
       \biggr),
\edm
as given in (\ref{V1NLO}). Numerically, this dependence is less
pronounced than in the resummed cross section. 
For $z\lesssim 0.1$, the growing scale dependence of the multiplicity
in the ${\cal O}(\alpha_s)$ calculation 
indicates that unaccounted higher-order effects become important and
are needed to improve the theory predictions.
For example, including the ${\cal O}(\alpha^2_s)$ coefficient 
${\cal C}_{ba}^{(2)}$ in the resummed calculation will
be necessary to improve the description of the charged particle multiplicity
in the small-$z$ region.

\section{Discussion and conclusions}
\label{SecConclusion}
In this paper we have continued the discussion of the effects of
multiple parton
radiation in semi-inclusive deep-inelastic scattering that was started in
\cite{nsy1999}. We have demonstrated that these effects
influence a large class of observables and  
can be described with the help of a
resummation formalism \cite{Meng2} 
that is similar to the resummation formalism 
used to improve the
theoretical description of energy-energy correlations in $e^+e^-$
hadroproduction \cite{CSepem}, and 
small-$p_T$ vector boson production at
hadron-hadron colliders \cite{CSS}.
 
Although 
this formalism needs further improvement, in particular in 
the procedure for 
matching the resummed curve with the perturbative result in the transition
region, 
it already improves the agreement between  
the theory and the data and provides interesting insights 
about qualitative features of semi-inclusive DIS. We have demonstrated
that the resummation formalism describes the
pseudorapidity distributions of the transverse energy flow more
accurately than the NLO calculation; this formalism also has 
good potential to improve the description of various distributions of 
particle multiplicity. 

From our analysis we conclude that the experimentally measured 
$q_T$ distribution for the energy flow broadens rapidly as $x \ra 0$.
This rapid broadening of the $q_T$ distribution can be
realized if the nonperturbative Sudakov factor in the resummed energy flow
increases as $1/x$. Similarly, the $q_T$ distribution for the 
charged particle multiplicity broadens rapidly when $z \ra 0$,
which is consistent with the nonperturbative Sudakov factor increasing as
$z^{-1.4}$. The sDIS nonperturbative Sudakov
factors at small values of $x$ and $z$ are therefore qualitatively different  
from the known nonperturbative Sudakov factors for vector boson
production and $e^+e^-$ hadroproduction, which do not depend on the
longitudinal variables at all. The rapid growth of the nonperturbative
Sudakov factor in sDIS might indicate that the $ep$ collider HERA
tests the resummation formalism in a new dynamical regime, which was not
studied yet at colliders of other types.

There are several theoretical aspects of the resummation formalism
that can be clarified when more experimental data are published.
Perhaps the largest uncertainty in the predictions of the resummation
formalism comes from the unknown nonperturbative contributions, which
in the $b$-space formalism  
are included in the nonperturbative Sudakov factor $S^{NP}(b)$. 
In this paper, we have presented simple parametrizations of $S^{NP}(b)$ 
for the transverse
energy flow (cf. Eq.\,(\ref{SNPz})) and charged particle multiplicity 
(cf. Eq.\,(\ref{SNPchgd})). These parametrizations were found by 
fitting the resummed energy flow and charged particle multiplicity 
to the data from Refs.\,\cite{H1z1,H1z2} and Ref.\,\cite{ZEUSchgd96},
respectively. Experimental measurements outside the range of
those data will make it possible to further improve these
parametrizations and, hence, the accuracy of the resummation formalism.
 
The most
straightforward way to study $S^{NP}(b)$ is
by measuring the variation of 
the $q_T$  spectra
of physical quantities due to variations of
one kinematical variable, with other variables fixed or varying only in a
small range. For the energy flow, it would be beneficial to obtain more
data at $x > 10^{-2}$, where the predictions of the resummation
formalism
can be tested more reliably, without potential uncertainties due to the 
small-$x$ physics. Another interesting question is the dependence of
the
nonperturbative Sudakov factor on the virtuality $Q$ of the vector boson. 
This dependence can be tested by studying the $q_T$ spectra in a
range of $Q$ with sufficient experimental resolution in the current
fragmentation region. Finally, to study effects of multiple parton
radiation
on semi-inclusive production of individual hadrons, it will be
interesting to see the $q_T$ spectra for particle
production multiplicities 
with the additional constraint $z > 0.05 \sim 0.1$, {\it i.e.,} in
the kinematical 
region where the parametrizations of the fragmentation functions
are known reasonably well.


\section*{Acknowledgements}
We would like to thank J. Collins, D. Soper, G. Sterman and other 
members of the CTEQ Collaboration for useful
discussions. 
The authors also thank 
S.~Kretzer for providing a Fortran code with the parametrizations of
the fragmentation functions. 
This work was supported under the NSF grant PHY-9802564.

\section*{Appendix A: Summary of notations}
Detailed explanations of the notations are provided in \cite{nsy1999}.

We limit the discussion to the case of the semi-inclusive DIS process
\begin{equation}
e+A\rightarrow e+B+X \label{sdis2}
\end{equation}
at the $ep$ collider HERA. Here
$e$ is an electron or positron, $A$ is a proton,
$B$ is a hadron observed in the final state, and $X$ represents
any other particles in the final state in the sense of inclusive
scattering.
We denote the momenta of $A$ and $B$ by $p_{A}^{\mu}$ and
$p_{B}^{\mu}$, and the momenta of the lepton in the initial
and final states by $l^{\mu}$ and $l^{\prime\mu}$.
Also, $q^{\mu}$ is the momentum transfer to the hadron system,
$q^{\mu}=l^{\mu}-{\lp}^{\mu}$.
We assume that the initial lepton and hadron interact only through a single
photon exchange.
Therefore, $q^{\mu}$ also has the
meaning of the 4-momentum of the exchanged virtual photon $\vg$.

The kinematics of the process (\ref{sdis2}) can be discussed
conveniently with the help of the following five Lorentz scalars:
\begin{itemize}
\item  the square of the center-of-mass energy $\sqrt{\SeA}$ of the
initial hadron and lepton,
\begin{equation} S_{eA}=(p_{A}+l)^{2}=2 p_{A}\cdot l;      \label{SeA}
\end{equation}
\item the momentum transfer from the leptonic to the hadronic part of
the process,
\begin{equation}
Q^{2}=-q^{2}=2\ell\cdot\ell^{\prime};
\end{equation}
\item the Bjorken variable,
\begin{equation}
x=\frac{Q^{2}}{2p_{A}\cdot{q}};
\label{x}
\end{equation}
\item  a variable $z$ describing energy transfer from the
composite photon-proton intermediate state to the observed final
hadron $B$,
\begin{equation}
\label{z}
z=\frac{p_{B}\cdot p_{A}}{q\cdot p_{A}}=\frac{2x p_{B}\cdot p_{A}}{Q^{2}};
\end{equation}
\item a scale $q_{T}$ that serves as a control parameter in the
resummation formalism,
\begin{equation}
q_{T}^{2}=-q_{t}^{\mu}q_{t\mu},
\end{equation}
where
\begin{equation}
q_{t}^{\mu}=q^{\mu}-p_{A}^{\mu}\frac{q\cdot p_{B}}{p_{A}\cdot p_{B}}
-p_{B}^{\mu}\frac{q\cdot p_{A}}{p_{A}\cdot p_{B}}. 
\end{equation}
\end{itemize}

Let $a$ denote the parton in $A$ that participates in the hard scattering,
with momentum $p_{a}^{\mu}=\xi_{a}p_{A}^{\mu}$.
Let $b$ denote the parton of which $B$ is a fragment,
with momentum $p_{b}^{\mu}=p_{B}^{\mu}/\xi_{b}$.
We use the symbol ``$\widehat{\quad}$'' to denote quantities at the parton level.
Correspondingly, we introduce the parton-level Lorentz scalars
$\xh,\,\zh$ that are analogous to the hadron-level scalars $x,\ z$:
\beq
\xh  = \frac{Q^{2}}{2p_{a}\cdot{q}}= \frac{x}{\xi_a}, 
\eeq
\beq	
\zh  = \frac{p_{b}\cdot p_{a}}{q\cdot p_{a}}=\frac{z}{\xi_b}.
\eeq

The paper concentrates on the discussion of 
experimental data presented in the hadronic center-of-mass (hCM) coordinate
system, 
which is defined to be the center-of-mass frame of
the proton $A$ and the virtual photon $\gamma^*$, with the $z$ axis
pointing in the direction of the photon's momentum. The momenta of $A$
and
$\vg$ in the hCM frame are
\bea
p_A^{\mu} &=& \left\{\frac{W^2+Q^2}{2 W},0,0,-\frac{W^2+Q^2}{2
W}\right\},\\
q^{\mu} &=& \left\{\frac{W^2-Q^2}{2 W},0,0,\frac{W^2+Q^2}{2
W}\right\},
\eea
where
\beq
W \equiv Q \sqrt{\frac{1}{x}-1}.
\eeq
The momentum of the
outgoing hadron $B$ in this frame lies in the $xz$ plane,
and its component perpendicular to the $z$ axis is
\beq   \label{pTB}
p_T = z q_T.
\eeq
The energy and longitudinal component of the momentum of $B$ are
\bea
E_B & = & z \frac{W^2 + q_T^2}{2 W},\\
p_{B}^z & = & z \frac{W^2 - q_T^2}{2 W}.
\eea

The pseudorapidity of $B$ in this  frame is  simply related to $q_T$,
by
\beq
\label{qTetacm}
 q_T = W e^{-\eta_{cm}}.
\eeq
The region of small $q_T$, where our resummation
calculation is relevant, 
corresponds to the region of large pseudorapidity
in the hCM frame.

We denote by $\varphi$ the angle between the $xz$ plane
(the plane of the momenta $\vec p_A$ and $\vec p_B$) and the plane
of the lepton momenta $\vec \ell$ and $\vec \ell^\prime$. 
In terms of the angle $\varphi$ and the boost parameter 
$\psi = \cosh^{-1} \left( 2x \SeA/Q^{2} - 1\right)$, 
the momenta of the leptons in the hCM frame are 
\bea
l^\mu &=& \Biggl\{\frac{1}{4 W}
\biggl((W^2+Q^2) \cosh \psi + W^2 - Q^2\biggr), \frac{Q}{2} \sinh\psi \cos \varphi,-\frac{Q}{2} \sinh\psi \sin \varphi, \nonumber \\
&&\frac{1}{4 W}
\biggl((W^2+Q^2) + (W^2 - Q^2)\cosh \psi \biggr)  \Biggr\};\\
l^{\prime \mu} &=& \Biggl\{\frac{1}{4 W}
\biggl((W^2 + Q^2) \cosh \psi - W^2 + Q^2\biggr), \frac{Q}{2}
\sinh\psi \cos \varphi,-\frac{Q}{2} \sinh\psi \sin \varphi, \nonumber
\\
&&\frac{1}{4 W} \biggl(-W^2 - Q^2 + (W^2 - Q^2)\cosh \psi \biggr)  \Biggr\}.
\eea
It is convenient to decompose the sDIS cross section into a sum 
over a complete set of functions $A_\rho (\psi, \varphi)$ of the azimuthal
angle $\varphi$ and the parameter $\psi$
 \cite{Meng1}:
\beq
\frac{d \sigma_{BA}}{d x d z  dQ^{2}  dq_T^2 d\varphi}=
\sum_{\rho} \prescr{\rho} V_{BA}( x , z, Q^2, q_T^2)
A_{\rho}(\psi,\varphi). \label{ang2}
\eeq
If the scattering of leptons on hadrons is caused by a
parity-conserving
interaction, then 
the spin-averaged sDIS cross section receives contributions only from 
four angular functions $A_\rho (\psi, \varphi) $, which are
\bea
 A_1 &=& 1+ \cosh^2 \psi,    \nonumber              \\
 A_2 &=& -2,                  \nonumber              \\
 A_3 &=& - \cos \varphi \sinh 2\psi, \nonumber \\
 A_4 &=& \cos 2 \varphi \sinh^2 \psi.
\eea
Since the angular functions $A_3$ and $A_4$ are proportional to 
$\cos \varphi$ and $\cos 2\varphi$, respectively, they do not
contribute to  the cross section that is   
integrated over the azimuthal angle $\varphi$. Hence, the $\varphi$-integrated
cross section $d\sigma/(dx dz dQ^2 dq_T^2)$ depends only  
on the structure functions $A_1$ and $A_2$.

\section*{Appendix B. Perturbative coefficients in the resummed
cross section}
The Wilson coefficient
functions ${\cal C}^{in}_{ba}$ and ${\cal C}^{out}_{ba}$, and the perturbative 
part $S^P(b,Q)$ of the Sudakov factor in (\ref{W}) 
can be expanded in powers of $\alpha_s$ \cite{Meng2,nsy1999}, as 
\bea
S^{P}(b , Q, C_1, C_2) &=& \int_{C_{1}^{2}/b^{2}}^{C_{2}^{2}Q^{2}}
\frac{d\ov \mu^{2}}{\ov \mu^{2}}\sum_{k=1}^{\infty} 
\Biggl(\frac{\alpha_s (\mu)}{\pi} \Biggr)^k
\biggl\{ A^{(k)}(\alpha_s(\ov \mu), C_1)
\ln\frac{C_{2}^{2}Q^{2}}{\ov \mu^{2}}                \nonumber \\
&+& B^{(k)} (\alpha_s(\ov \mu ), C_1,C_2)\biggr\}; \label{sud_p} \\
{\cal C}^{in}_{ba}(\xh, C_1, C_2, b, \mu_F) &=&
\sum_{k=0}^{\infty} {\cal C}^{in
(k)}_{ba}(\xh, \mu_F b)
\Biggl(\frac{\alpha_s (\mu_F)}{\pi} \Biggr)^k;
                                                              \\
{\cal C}^{out}_{ba}(\zh, C_1,C_2, b, \mu_F) &=&
\sum_{k=0}^{\infty} {\cal C}^{out
(k)}_{ba}(\zh, \mu_F b) \Biggl(\frac{\alpha_s (\mu_F)}{\pi}\Biggr)^k.
\label{cout0}
\eea
Up to the order ${\cal O}(\alpha_s)$, the coefficients in 
Eqs.\,(\ref{sud_p})-(\ref{cout0}) are listed below. 
\begin{itemize}
\item For the Sudakov factor,
\bea
A^{(1)} &=& C_F, \\
B^{(1)} &=& 2 C_F \ln \biggl(\frac{e^{-3/4} C_1}{b_0 C_2}\biggr).
\eea
Here $C_F=4/3$ is a color factor; $\mu_F$ is a factorization scale;
$b_0 = 2 e^{-\gamma_E}$; $C_1$ and
$C_2$ are constants that are conventionally chosen such that 
$C_1/C_2 = b_0$. 
\item For the coefficient functions,
\bea
{\cal C}^{in (0)}_{ij} (\xi, \mu_F b) &= &  
{\cal C}^{out (0)}_{ij} (\xi, \mu_F b) =
\delta_{ij} \delta (1-\xi), \\
     {\cal C}^{in (0)}_{ag} &=& {\cal C}^{out (0)}_{ga}=0,\\
{\cal C}^{in (1)}_{ij}(\widehat{x},\mu_F b)&=& \frac{C_F}{2}(1-\xh) -
P_{qq}(\xh) \ln \Bigl(\frac{\mu_F b} {b_0}\Bigr)  \nonumber\\
&-& C_F \delta(1-\xh) \biggl( \frac{23}{16}+ \ln^2 \Bigl(
\frac{e^{-3/4} C_1}{ b_0 C_2} \Bigr) \biggr),\label{C1in}
\\
{\cal C}^{in (1)}_{jg}(\widehat{x},\mu_F b)  &=& \frac{1}{2}\xh (1-\xh) -
P_{q g} (\xh) \ln \Bigl(\frac{\mu_F b}{b_0}\Bigr),\label{C1in2}
\\
{\cal C}^{out (1)}_{ij}(\widehat{z},\mu_F b) &=& \frac{C_F}{2}(1-\zh) -
P_{qq}(\zh) \ln \Bigl(\frac{\mu_F b} {b_0 \zh}\Bigr) \nonumber \\
&-& C_F \delta(1-\zh) \biggl( \frac{23}{16}+ \ln^2 \Bigl(
\frac{e^{-3/4} C_1}{ b_0 C_2} \Bigr) \biggr),\label{C1out}
\\
{\cal C}^{out (1)}_{gj}(\widehat{z},\mu_F b) &=& \frac{C_F}{2}\zh -
P_{gq} (\zh )\ln \Bigl(\frac{\mu_F b}{b_0 \zh}\Bigr), \label{C1out2}\\
{\cal C}^{in (1)}_{gg} &=& {\cal C}^{out (1)}_{gg} = 0.
\eea
Here $P_{qq}$ and $P_{qg}$ are one-loop splitting functions \cite{DGLAP}.
The subscripts $a,b$ correspond to quarks, antiquarks and gluons; the
subscripts $i,j$ correspond to the quarks and antiquarks only.
\end{itemize}
 
\tighten

\setlength{\unitlength}{1cm}
\begin{figure}[ht]
\begin{picture}(10,20)
\put(3,11){
\epsfysize 9.5 cm
\epsffile{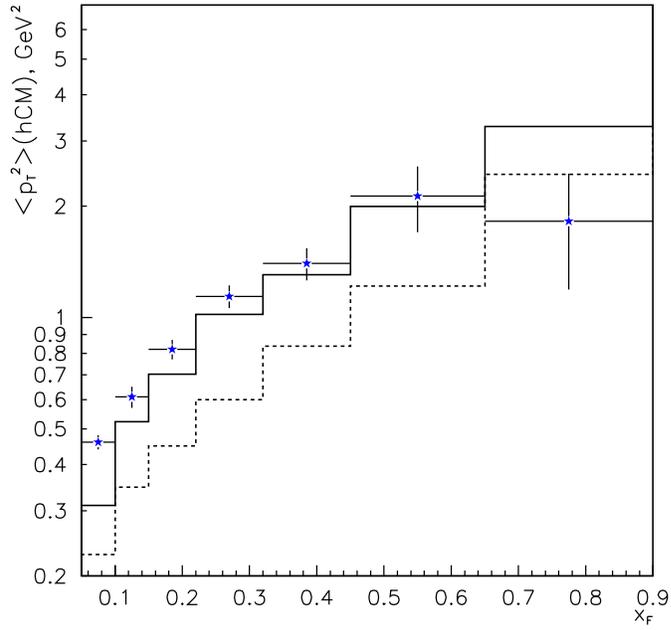}}
\put(3,1){
\epsfysize 9.5cm
\epsffile{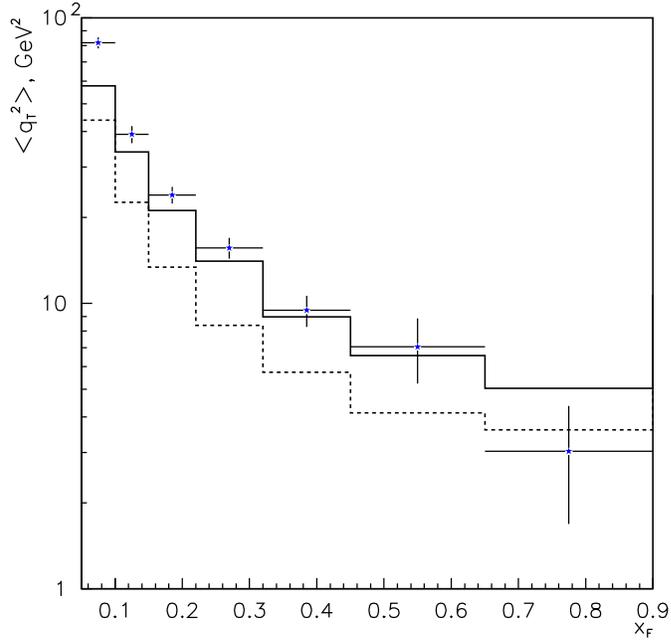}
}
\put(8,10.5){\makebox{(a)}}
\put(8,0.5){\makebox{(b)}}
\end{picture}
\caption{\label{seagull}
The distributions (a) $\langle p_T^2 \rangle$~vs.~$x_F$
and (b) $\langle q_T^2 \rangle$~vs.~$x_F$ for the charged particle
multiplicity
at $\langle W \rangle = 120 \mbox{ GeV}, \ 
\langle Q^2 \rangle = 28 \mbox{ GeV}^2$. The
experimental points for the distribution 
$\langle p_T^2 \rangle$~vs.~$x_F$ 
are from Fig.\ 3c of Ref.\,\protect\cite{ZEUSchgd96}. 
The ``experimental'' points for the distribution
$\langle q_T^2 \rangle$~vs.~$x_F$ are derived using Eq.\,(\ref{pT2toqT2}). 
The solid and dashed curves
correspond to the resummed and the next-to-leading order ($\mu_F = Q$) 
multiplicity, respectively.}
\end{figure}

\end{document}